\documentclass[twocolumn,aps,prl,showpacs,reprint]{revtex4-1}

\usepackage{times}
\usepackage{amsmath,amssymb}           
\usepackage{graphicx}                   
\usepackage{subfigure}                  

\begin{document}

\title{Resonant Metalenses for Breaking the Diffraction Barrier}
\author{Fabrice Lemoult}
\author{Geoffroy Lerosey}
\email[]{geoffroy.lerosey@espci.fr}
\author{Julien {de Rosny}}
\author{Mathias Fink}
\affiliation{Institut Langevin, ESPCI ParisTech \& CNRS, Laboratoire Ondes et Acoustique, 10 rue Vauquelin, 75231 Paris Cedex 05, France}

\date{14 April 2010}

\begin{abstract}
We introduce the resonant metalens, a cluster of coupled subwavelength resonators. Dispersion allows the conversion of subwavelength wavefields into temporal signatures while the Purcell effect permits an efficient radiation of this information in the far-field. The study of an array of resonant wires using microwaves provides a physical understanding of the underlying mechanism. We experimentally demonstrate imaging and focusing from the far-field with resolutions far below the diffraction limit. This concept is realizable at any frequency where subwavelength resonators can be designed. 
\end{abstract}

\pacs{41.20.-q, 81.05.Xj, 78.67.Pt}


\maketitle

Within all areas of wave physics it is commonly believed that a subwavelength wavefield  cannot propagate in the far field. This restriction arises from the fact that details with physical dimensions much smaller than the wavelength are carried by waves whose phase velocity exceeds that of light in free space, which forbids their propagation. Such waves, usually referred to as evanescent waves, possess an exponentially decreasing amplitude from the surface of an object \cite{goodman}.

 	Numerous works have been devoted to overcome this diffraction limit, starting from the early 20th century and the proposal by Synge of the first near-field imaging method \cite{Synge}. Since this seminal work, near-field microscopes have been demonstrated from radio frequencies up to optical wavelengths, achieving resolutions well below the diffraction limit \cite{pohl,lewis,betzig,Zenhausern,taubner,vlahacos}. Fluorescence based imaging methods have also been proposed, which allow deep subwavelength imaging of living tissues \cite{Hell}.
 	 Such concepts, however, employ several measurement of the same sample in order to beat the diffraction limit through image reconstruction procedures. Finally, various new concepts have been proposed such as far-field superlens and hyperlens \cite{JOSADurant2006,Liu,Pendry}, demonstrating moderate sub-diffraction imaging down to a quarter of the optical wavelength. 
 	
In this Letter, we introduce the concept of resonant metalens, a lens composed of strongly coupled subwavelength resonators, and prove that it permits sub-diffraction imaging and focusing from the far-field using a single illumination. Our concept resides in exploiting time coded far-field signals for spacial sub wavelength resolution \cite{Grbic,Stockman}. Studying the specific case of an array of resonant wires, we explain theoretically, prove numerically, and demonstrate experimentally how this lens converts the subwavelength spatial profile of an object into a  temporal signature and allow efficient propagation of this information towards the far-field. We achieve far-field imaging and focusing experiments with resolutions of respectively  $\lambda/80$ and  $\lambda/25$, well below the diffraction limit.

	The notion of evanescent waves finds its roots in a mathematical formalism which perfectly fits that of infinite interfaces. Indeed, the projection of an infinitely extended subwavelength varying field onto a basis of free-space radiations results in a null value: the field sticks to the object, the waves are evanescent. However, this does not hold anymore when considering objects of finite dimensions. In these cases the sub-diffraction details of the latter contribute to the far-field due to finite size effects. Practically, the efficiency of this conversion decreases dramatically with the diminishing size of the subwavelength spatial variation. Hence, any measurement of sub-diffraction details in the far-field appears very tedious since their contribution to the total field is much weaker than those of propagating diffraction-limited waves.
	
	Such a monochromatic approach seems limited since all spatial information (subwavelength or not) propagating away from an object mixes in a unique wavefield. Another difficulty is that the smaller the detail to be resolved, the weaker its contribution to the radiation of the object. The solution to both issues lies in the concept of resonant metalens. We define the latter as a cluster of resonators arranged on a subwavelength scale forming a lens in the near-field of an object, and illuminated with broadband wavefields. Its mechanism, which will be developed in detail later for a special case made out of conducting wires, can be explained intuitively.

	On one hand, placing $N$ identical resonators on a subwavelength scale introduces a strong coupling between them, which splits the original resonance frequency into a band of $N$ different ones, analogous to Kronig-Penney potential wells in solid state physics \cite{kronig}. The cluster of oscillators can be described by a set of $N$ eigenmodes and eigenfrequencies, the latter being distributed on an interval which depends on the strength of the coupling. Illuminated by a broad range of energies, the near-field on any
subwavelength object, at the resonant metalens input, decomposes onto the modes of the system with a unique set of phases and amplitudes. Since all those modes are excited at a different frequency (ignoring any degeneracy due to symmetry), the information of the object gets translated in the spectrum of the field generated in the lens. In the temporal domain, harnessing the modal dispersion of the resonant metalens permits the conversion of the subwavelength details of the object into a temporal signature.

\begin{figure*}
	\begin{center}
		\includegraphics[width=13.8cm]{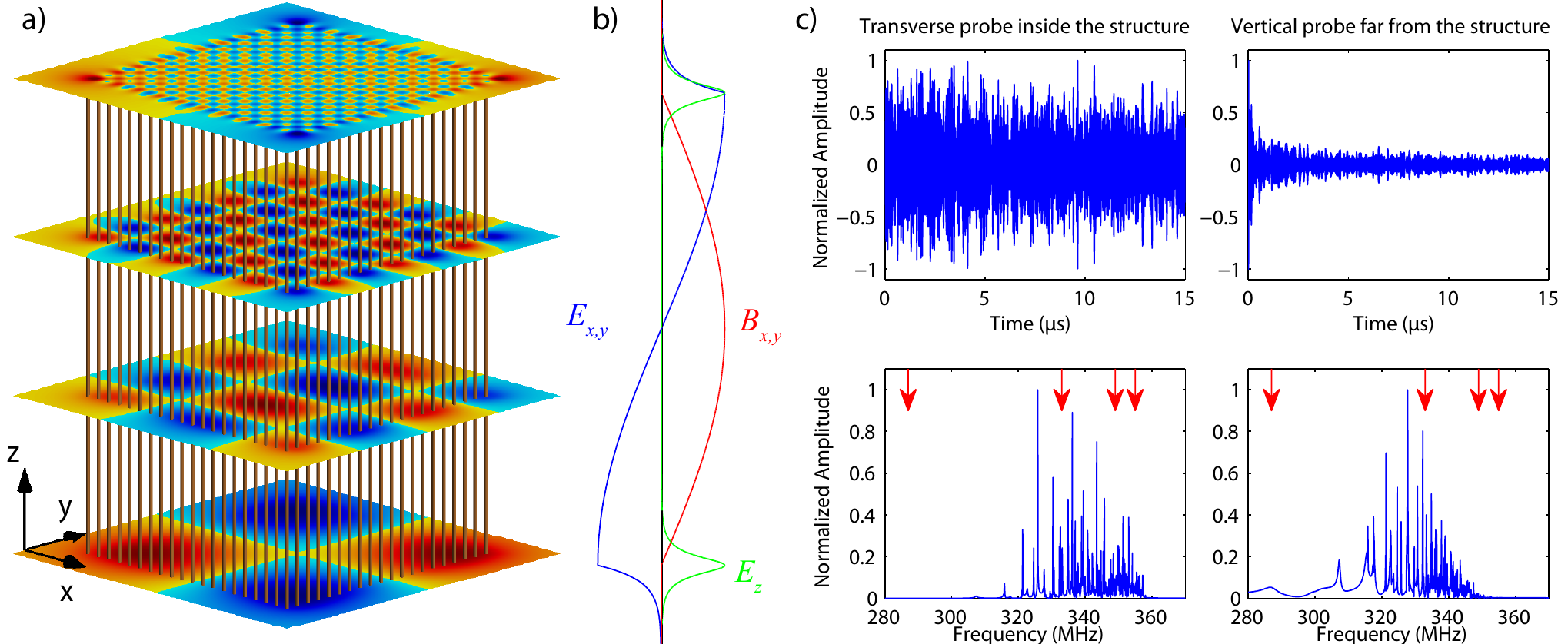} 		
	\end{center}
	\caption{\label{figure1} (a) Three dimensional representation of the medium described in the text. Superimposed: amplitude of $E_x$ TEM Bloch modes (1,1), (2,3), (5,6) and (19,19). (b) Longitudinal profile of electric field $E_{xy}$ (blue), $E_z$ (green) and magnetic field $B_{xy}$ (red). (c) Results of the transient 3D simulations: inside the structure and in the far-field. Red arrows: resonance frequencies of the 4 modes mapped in (a).}
\end{figure*}

	On the other hand, our lens grants an efficient propagation of subwavelength information towards the far-field. Ignoring the intrinsic losses of the material, the energy stored in a given mode of the lens be dissipated through radiative decay only. Any eigenmode of the system can be defined by a wavevector, which will be referred to as a transverse wavevector $\roarrow{k_\perp}$, and the higher its norm ${k_\perp}$, the lower the efficiency of the conversion from this mode to propagating waves. This inefficiency enhances the lifetime of the mode in the structure. Due to the Purcell effect \cite{purcell1946}, the higher the lifetime of a given mode, the better the coupling from the illuminated object to this mode: surprisingly this counterbalances the effect of the weak coupling of the mode to the far-field. Ignoring the intrinsic losses and due to the resonant nature of the lens, every subwavelength mode, independently of $k_\perp$, radiates in the far-field an equivalent amount of energy over time. Because of their higher lifetimes, deeper subwavelength modes tend to escape the lens tardily. The intrinsic losses diminish the lifetimes of the modes proportionally to their localization on the resonators (the higher $k_\perp$, the shorter the lifetime): this, in turn, will limit the resolution of the resonant metalens.
		
	We focus now on the peculiar resonant metalens consisting of an ordered collection of parallel conducting wires. Interestingly, a wire array forms in the transverse plane a subwavelength arrangement of resonators thanks to the resonance occurring along the longitudinal dimension. We numerically study a medium (Fig. \ref{figure1}.a) made out of a square periodic lattice of $N$$ \times$$ N$ ($N$=$20$) perfect electric conductor wires of diameter $d$ ($3 \textrm{ mm}$), with equal length of $L$ ($40 \textrm{ cm}$), and a period of $a$ ($1.2  \textrm{ cm}$) between the wires in both transverse directions ($xy$ plane). The array lies in air and thus the first resonance frequency of a single wire occurs if its length matches half a wavelength ($f_0$=$375 \textrm{ MHz}$). At this frequency, the spacing between the resonators corresponds roughly to  $\lambda/70$, meaning that the wires are strongly coupled. Fortunately, this lens can be analyzed using the theory of the "wire media" \cite{belov,shvets} instead of calculating all of the coupling coefficients. The field propagating in the structure can be expanded in Bloch modes because of the periodic nature of the medium, and inside the structure \cite{shvets}, the boundary conditions and the deep subwavelength period impose a transverse electromagnetic nature of the field (TEM, $E_z$=$B_z$=0). Due to the transverse finiteness of the system, the $\roarrow{k_\perp}$ are quantified, $\roarrow{k_\perp}$=$\frac{\pi}{D}(m.\roarrow{e_x}$+$n.\roarrow{e_y})$, with integers $(m,n)$$\in$$[\![1;N]\!]^2$ and $D$ the size of the medium in the $x$ and $y$ dimensions, $D$=$a(N$-$1)$.
	
	
	Those modes present a constant longitudinal wavevector $k_z$ independent of $\roarrow{k_\perp}$: the longitudinal propagation is dispersionless and the phase velocity equals that of plane waves in the host matrix \cite{belov,shvets}. The resonant behavior of the modes can be revisited at the light of the TEM approach: due to their finite length, the wires constitute Fabry-Perot cavities for the modes \cite{belov}. This specific kind of resonant metalens presents the great advantage of being analyzable within two different and complementary frames: the subwavelength coupled oscillators and the Fabry-Perot like TEM Bloch modes.
	
	We have performed numerical simulations whose details are presented in \cite{SI}. The structure is excited with a small electric dipole $2 \textrm{ mm}$ away from the lower interface, which we define as the input of this resonant metalens. The emitted signal is a $5 \textrm{ ns}$ pulse centered around $300 \textrm{ MHz}$. As expected, the fields are transverse electromagnetic (Fig. \ref{figure1}.b). We underline here that taking advantage of the Fabry-Perot resonance permits an efficient electrical coupling to the eigenmodes since the electric field is maximum at both ends of the wire medium (Fig. \ref{figure1}.b). Naturally, the field generated by the source expands on the eigenmodes, and in figure \ref{figure1}.a, superimposed on the structure, we map the electric field for four different modes chosen among the $N^2$ ones. The time varying fields in the structure and in the far-field are plotted alongside their spectra in Figure \ref{figure1}.c. 
In fact, the unique decomposition of the source onto the eigenmodes manifests itself in the spectrum and the temporal evolution of the near-field. Probing now the far-field in the $(xy)$ plane, which is vertically polarized since $E_x$ and $E_y$ are odd while the boundary conditions impose an even $E_z$ (Fig. \ref{figure1}.b), the spectrum looks very similar to that of the near-field. This proves that the near-field converts efficiently to propagating waves, as predicted. Thanks to the modal dispersion of the lens, subwavelength details ranging from $k_0$ to $30 \textrm{ } k_0$ ($\lambda$/a) decompose onto the modes, are converted into temporal information and the profile of the source propagates toward the far-field stored in the spectrum of the field. To exemplify this, we point with arrows the exact resonance frequencies of the 4 modes mapped in Fig. \ref{figure1}.a. on the near-field and far-field spectra. 
		
The dispersion of the wire medium, a key issue for our resonant metalens, can be derived and understood quite easily: even though TEM Bloch modes share the same phase velocity, their penetration depth in air at the $z$=$-L/2$ and $z$=$L/2$ interfaces depends on their transverse wavevector $k_\perp$. Consequently, the longitudinal extension of the mode depends on $k_\perp$, which modifies the effective length of the Fabry-Perot cavity, and gives the following dispersion relation \cite{More}:
	
\begin{equation}
\frac{f_0}{f}=1+\frac{2}{\pi\sqrt{(k_\perp/k_0)^2-(f/f_0)^2}}
\end{equation}

\begin{figure}
\begin{center}
		\includegraphics[width=7.3cm]{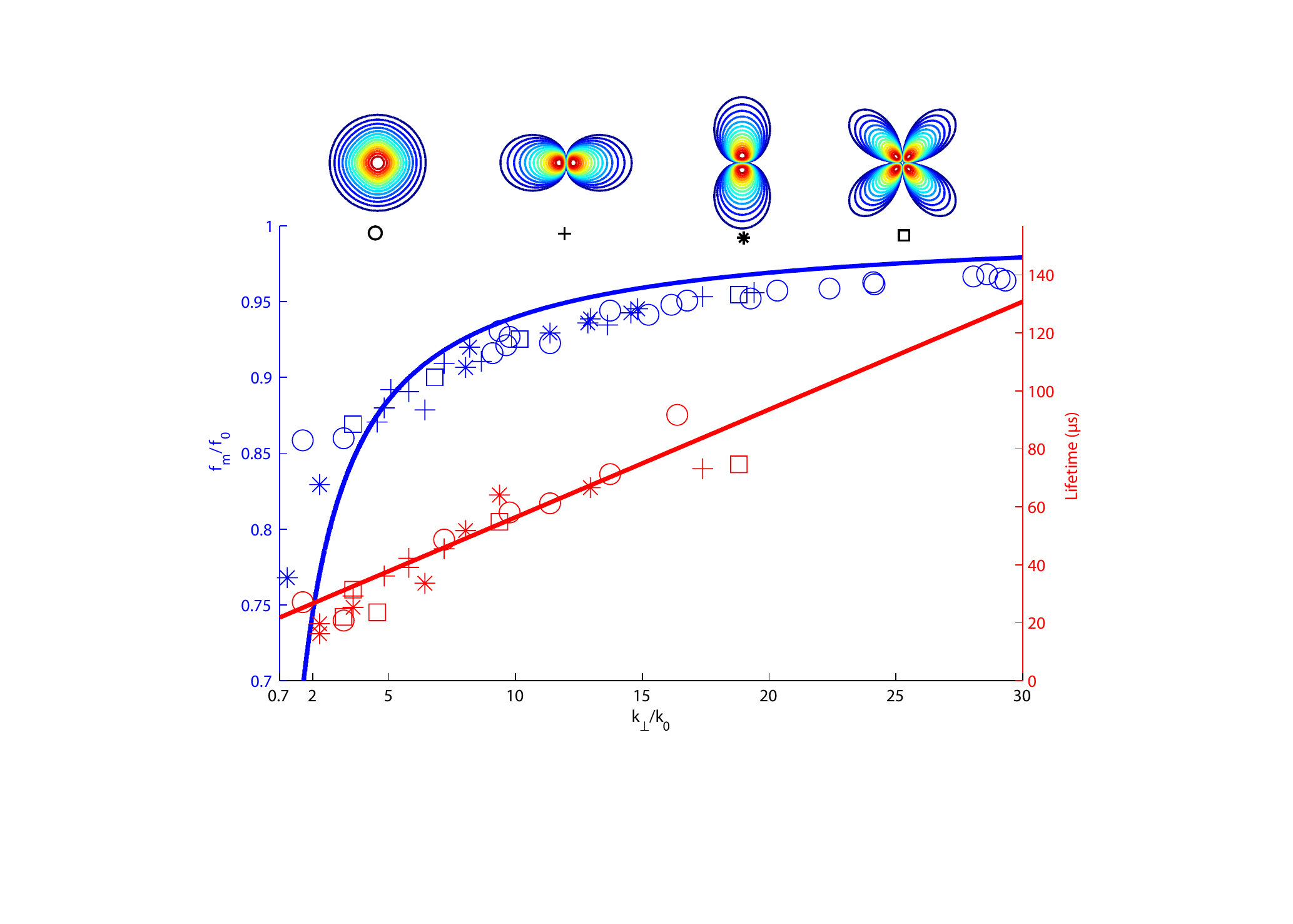} 		
	\end{center}
	\caption{\label{figure2} The dispersion relation (solid blue line) in terms of the modal resonance frequencies $f_m$ (normalized to $f_0$) versus ${k_\perp}$ (normalized to $k_0$). Symbols: dispersion relation extracted from the simulations (the 4 symbols represent the 4 radiation patterns). Red symbols: lifetimes extracted from simulations as a function of ${k_\perp}$ (normalized to $k_0$). Red solid curve: linear fit of slope $(D{k_\perp})^{-1}$.}
\end{figure}

This theoretical dispersion relation is plotted in terms of the modes resonance frequency $f_m$ versus $k_\perp$ (both normalized to the original Fabry-Perot resonance frequency $f_0$ and wave number $k_0$) in Figure \ref{figure2}. We also extracted from the simulations the modes and calculated their transverse wavenumber ${k_\perp}$. Those data are superposed with the theoretical curve, and the good agreement proves the validity of the model.

Equally interesting, the far-field propagation of the subwavelength modes deserves explanation. The spectra in Figure \ref{figure1}.c show a diminution of the linewidth of the modes for the high transverse wavenumbers, consistently with the resonant metalens intuitive description we gave: the more subwavelength a wave mode, the higher its lifetime in the structure. To estimate the lifetime theoretically, we evaluate the efficiency of the conversion of the subwavelength modes to free-space waves \cite{More}:
 it is roughly proportional to  $(D{k_\perp})^{-1}$. 
The Purcell effect results in an increase of the coupling between the source (our "object") and the high ${k_\perp}$  modes of the resonant metalens. The measure of the source's return loss demonstrates the impedance matching in the presence of the metalens \cite{More}. 
The resonant nature of the modes matches the impedance of the small electric dipole, or equivalently, the Purcell effect compensates for the weak radiation of the deep subwavelength modes. 

We estimate the lifetimes of the modes in the metalens resulting from our simulations, through a time/frequency analysis. As we stopped the simulation after $15 \textrm{ $\mu$s}$ for calculation time issues, the lifetimes of the high ${k_\perp}$ could not be extracted. We plot the data in Figure \ref{figure2}, as well as a linear fit. An important remark concerns the harmony between the lifetimes of the modes and the corresponding dispersion relation. Indeed, the coding of subwavelength information in time requires that a maximum of the modes can be resolved. Here, nicely enough, the increase of the lifetime for modes of high ${k_\perp}$ counterbalances the flattening of the dispersion relation.

In Fig. \ref{figure2}, the four types of radiation pattern generated by this structure are also presented. Depending on the $m$ and $n$ indexes of $\roarrow{k_\perp}$, monopolar, dipolar ($x$ or $y$ oriented) or quadrupolar patterns coexist \cite{More}.
 This underlines that the lens, although subwavelength ($\lambda/3$), also possesses four spatial degrees of freedom, representing as many information channels exploitable for imaging and focusing \cite{PRLLemoult2009}. 

\begin{figure*}
	\begin{center}
		\includegraphics[width=13.7cm]{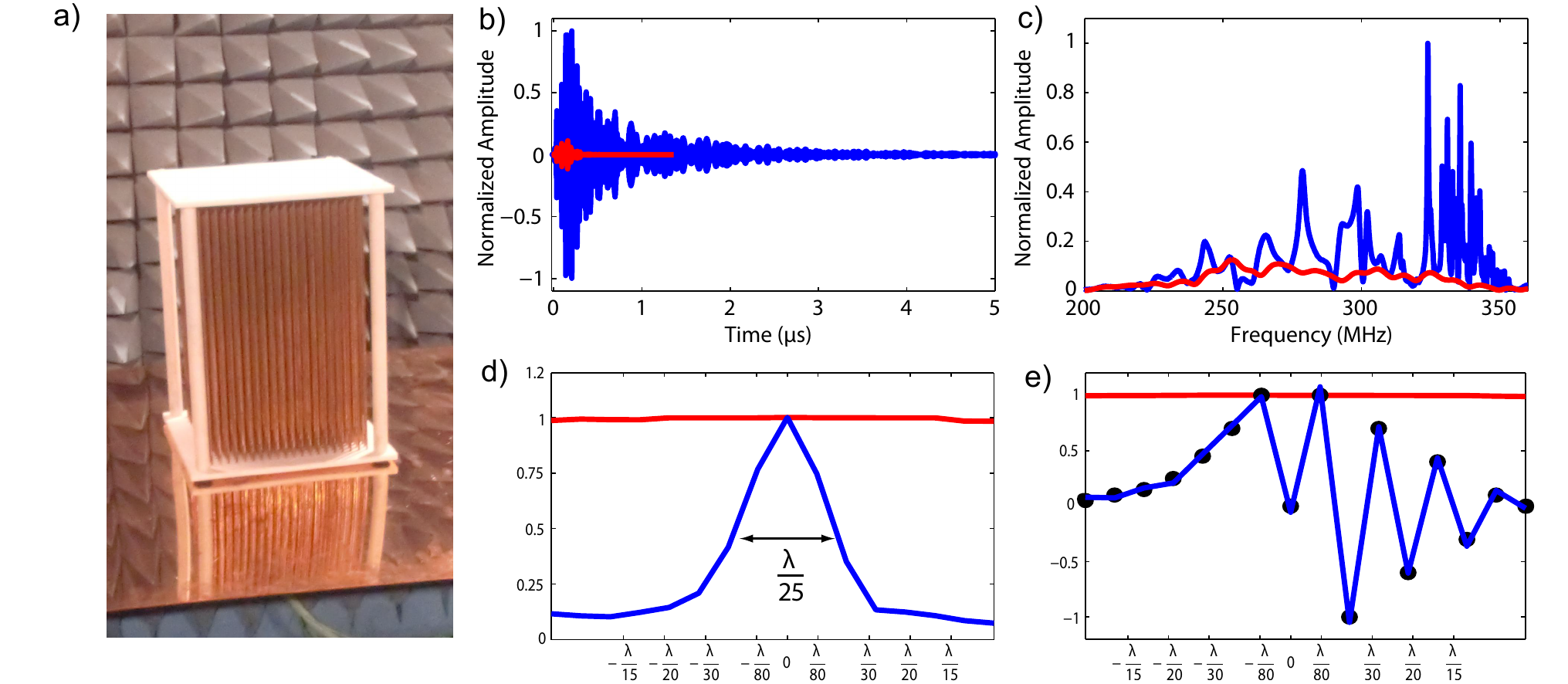} 		
	\end{center}
	\caption{\label{figure3} (a) The experimental resonant metalens on the ground copper plane. Experiments are performed in an anechoic chamber. (b-c) Signals and spectra received in the far-field after emission from central monopole with the lens (blue) and without as a control curve (red).  (d) Focal spot obtained after one channel Time Reversal of (b) from the far-field: a  $\lambda/25$ width is demonstrated in the presence of the resonant metalens (blue), no focusing without the lens (red). (e) An imaging experiment. 16 monopoles generate a subwavelength phase and amplitude profile in the near field of the lens (black points). The far-field is acquired on 8 antennas. We plot the result of the image reconstruction: a true  $\lambda/80$ resolved image of the initial pattern is reconstructed in the presence of the resonant metalens (blue) while it is impossible without (red).}
\end{figure*}

In order to seal the validity of the concept for real materials, we performed an experimental verification of the lens. The realistic lens presented in Figure \ref{figure3}.a replicates the simulated one, except that we use copper wires and a Teflon support ($\varepsilon_T$=$2.2$). Also, to avoid radiation leakage on the source cable as well as parasitic effects, we screen it by placing the structure $5 \textrm{ mm}$ on top of a 1 meter square ground plane. We measure in 8 directions the far-field generated by a small electric monopole located between the ground plane and the resonant metalens in an anechoic chamber \cite{SI}. The signal and spectrum for one direction, plotted in Figure \ref{figure3}.b.c, proves a very good agreement with the simulation results. Two remarks arise; First, the resonance frequencies show a redshift due to the Teflon structure. Second, the signals spread over a shorter time due to both the skin effect on the copper wires, whose permittivity is finite, and the losses in the Teflon; this effect manifests itself more clearly at high frequencies. Indeed, the higher  ${k_\perp}$, the more localized on the wires the modes, which decreases their lifetime due to ohmic losses. 

In Figure \ref{figure3}.d, we plot the result of a time reversal focusing experiment like in \cite{Lerosey2007,Lerosey2004} achieved from the far-field, in an anechoic chamber in order to measure the effect of the lens only. The focal spot obtained plotted alongside the control experiment (without the resonant metalens, no focusing at all) is roughly  $\lambda/25$ wide (at the central frequency of the excitation pulse). This means that subwavelength information of the sources have been converted in the far-field, and vice versa. 
Using time reversal, the green function between a source and the far-field antenna is "flipped" in time and reemitted. At each frequency, the signal is phase-conjugated, meaning here that all of the TEM Bloch modes generated in the lens add up in phase at a deterministic time, hence allowing the Time Reversal focusing. Since the decomposition of a point-like source onto the eigenmodes of the resonant metalens is unique, the modes add up incoherently at other positions. We point out that our precedent results \cite{Lerosey2007} can be interpreted at the light of the resonant metalens concept. We note that losses limit our focal spot sizes to  $\lambda/25$, but using other focusing techniques may shrink the spots even further.

Finally, we prove the imaging capabilities of the resonant metalens through a simple experiment: a subwavelength profile is generated at the input of the lens using simultaneously 16 monopoles \cite{SI}, and the far-field recorded in the anechoic chamber.
An inversion procedure with predesigned filters \cite{More}
 is used to reconstruct the profile, using the knowledge of each monopole temporal signature (Fig. \ref{figure3}.e). The subwavelength profile is perfectly reconstructed and an imaging resolution of about  $\lambda/80$ is demonstrated through this basic experiment. 

To conclude, this specific studied lens is scalable towards near-IR and in this range, the losses will increase, limiting the resolution. Using gain media in the matrix may counter this problem. More generally, we are currently working on a criterion linking the resolution achievable to the losses and typical size of the metalens. This lens presents degenerated modes because of the symmetry: adding some disorder in the spatial or resonant frequency distribution of the resonators, as well as in the matrix should lift this degeneracy and enhance dispersion. Finally, we would like to emphasize that the concept of resonant metalens should be realizable in any part of the electromagnetic spectrum, with any subwavelength resonator, such as split-rings \cite{Smith}, nanoparticles \cite{halas}, resonant wires \cite{muskens}, and even bubbles in acoustics \cite{leighton}.

We thank David F.P. Pile for his help with writing the manuscript and A. Souilah for the fabrication of the experimental prototype. F. Lemoult acknowledges funding from French "Direction G\'en\'erale de l'Armement".

\bibliography{biblio}

\end{document}